\documentclass[aps,prd,twocolumn,nofootinbib]{revtex4-2}
\usepackage{graphicx}
\usepackage{amsmath}
\usepackage[colorlinks=true]{hyperref}

\begin{document}
\title{Electronic structure calculation for superheavy elements Livermorium (Lv, Z=116) and Tennessine (Ts, Z=117) and their lighter analogs Te, I, Po, and At.}

\author{V. A. Dzuba}\email{v.dzuba@unsw.edu.au}
\author{V. V. Flambaum}\email{v.flambaum@unsw.edu.au}
\author{G. K. Vong}\email{g.vong@unsw.edu.au}
\address{School of Physics, University of New South Wales, Sydney 2052, Australia}

\begin{abstract}
Advanced theoretical techniques that combine the linearized coupled-cluster method, configuration interaction method, and perturbation theory are used to calculate energy levels, ionization potentials, electron affinities, field isotope shift, and static dipole polarizabilities of the superheavy elements Lv and Ts, along with their lighter analogs Te, I, Po, and At. Calculations for the heavy elements, Po, At, Lv, and Ts are used to address the gaps in the experimental data. Calculations for the lighter elements, Te and I (and partly Po and At) are used to demonstrate the accuracy of the calculations.
\end{abstract}

\maketitle

\section{Introduction}
The study of superheavy elements (SHEs), with atomic numbers $Z>104$, provides valuable insights into the behavior of electronic structure under extreme relativistic effects and strong electron correlations (see, e.g., recent reviews~\cite{smits2023,smits2024,ackermann2024,ye2025}). 
SHEs are particularly significant in the search for the predicted "island of stability," where nuclear shell models suggest the existence of longer-lived isotopes~\cite{ackermann2024,DFW17-IS}. However, despite experimental advances in SHE production, direct measurements remain exceedingly challenging because of the short half-lives and low production yields of these elements. 
Currently, there are no experimental data for the spectroscopic properties of all SHEs with $Z>102$. Numerous theoretical works have attempted to address this lack of experimental data~\cite{smits2023,smits2024}.  
The SHEs Livermorium (Lv, $Z=116$) and Tennessine (Ts, $Z=117$) remain the least studied.The reason is probably due to their complicated electronic structures with many valence electrons. Ref.~\cite{Borschevsky_2015} presents advanced calculations of the ionization potentials (IP) and electron affinities of the superheavy elements Mc, Lv, and Ts and their lighter analogs Bi, Po, and At.
Ref.~\cite{Z117} presents some transition energies and the polarizability of Ts.
However, data on the spectra of Lv and Ts remain very limited.
Note that knowing the spectra of an atom, together with the field isotope shift (FIS) parameters, is particularly important for the search for the island of stability in astrophysical data~\cite{DFW17-IS}. 

The significant challenge of accurate theoretical predictions for SHEs is due to the need for computational methods capable of treating strong electron correlation and relativistic effects in a self-consistent manner. The complexity of these systems renders many conventional methods computationally prohibitive. In this work, we employ a hybrid approach that combines the coupled-cluster single-double (SD)~\cite{Saf-CC,Dzu-CI-SD14} method with configuration interaction perturbation theory (CIPT)~\cite{CIPT} to calculate the atomic spectra, electron affinities, ionization potentials, field isotope shift, and polarizabilities of Lv and Ts. The CIPT method has previously been used in the calculation of other SHEs \cite{Db18,E118,SHE6d,E110-112}. 

The SD method is used to capture dominant core-core and core-valence correlation effects, while the CIPT method efficiently treats valence-valence correlations.
By integrating these two approaches, we achieve a balance between accuracy and computational feasibility, providing relatively precise predictions for electronic properties. This offers valuable theoretical benchmarks in the absence of experimental data.

\section{Method of calculations}

All atoms considered in the present work have an open $p$-shell. The configuration of the outermost electrons is $ns^2np^4$ for Te, Po, and Lv and $ns^2np^5$ for I, At, and Ts ($n=5$ for Te and I; $n=6$ for Po and At; $n=7$ for Lv and Ts). Sufficient experimental spectroscopic data are available only for Te and I~\cite{NIST_ASD}. Analysis of the NIST data for these atoms shows that there are no states with excitations from the $ns^2$ subshell {at least up to the energies of $\sim$~70000~cm$^{-1}$. Since we are mostly interested  in low-energy states, it is safe to assume} that the $ns$ electrons can be attributed to the core and only the $p$ electrons in the open shell are treated as valence electrons.
We use the so-called $V^{N-M}$ approximation~\cite{Dzu05} to perform the calculation ($N$ is the total number of electrons, $M$ is the number of valence electrons; $M=4$ for Te, Po, and Lv and $M=5$ for I, At, and Ts). The method involves solving the linearized coupled-cluster single-double (SD) equations for the core and valence states and using the configuration interaction (CI) technique for valence electrons.

Calculations start from the relativistic Hartree-Fock (RHF) procedure for the closed-shell core.
The RHF Hamiltonian can be written as 
\begin{equation}
    \hat{H}^{\mathrm{RHF}} =  c \boldsymbol{\alpha} \cdot \hat{\mathbf{p}} + (\beta - 1) mc^2  + V_{\rm nuc} + V^{N-M}.
    \label{h1}
\end{equation}
 Here $V_{\rm nuc}$ is nuclear potential obtained by integrating  the Fermi distribution of the nuclear charge  with the nuclear radius $R_N$
($V_{\rm nuc} = - \frac{Ze^2}{r}$ at $r \gg R_N$.)
$V^{N-M}$ is the self-consistent RHF potential for the closed-shell core after removing all valence electrons.

 {Using the $V^{N-M}$ approximation for systems with a relatively large number of valence electrons (four or five in the present work) may not seem ideal, as a poor initial approximation can lead to reduced accuracy. However, electrostatic potential produced by  valence electrons  in the area where the core electrons are located  is close to a constant potential. A constant potential does not affect wave functions.    Therefore, valence electrons have little effect on the core electron wave functions ~\cite{Dzu05}.   The $V^{N-M}$ approximation is poor only for the valence states. This deficiency is compensated by the completeness of the basis set. Most importantly, the key advantage of this approximation is that it provides a simple and consistent way to include the core-valence correlations. In particular, there  are no subtraction diagrams which give a very large contribution in the case of  the $V^{N}$ approximation and significantly complicate convergence of the diagram series.  The requirement for the $V^{N-M}$ approach is that the potential of the electronic core used in the CI calculations must be the same as the electron potential in the RHF calculations. See Ref.~\cite{Dzu05} for a detailed discussion.}

After the RHF procedure is completed, the single-electron basis states are calculated in the frozen core potential using the B-spline technique \cite{B-splines}.
The basis states are constructed as linear combinations of B-splines, found by diagonalizing the matrix of the RHF Hamiltonian in the B-spline basis. We calculate 40 B-spline states of order 9 in a box of radius $40 a_B$ {($a_B$ is Bohr radius)} in each partial wave up to $l_{\text{max}}= 6$. This choice of parameters ensures that the basis is sufficiently saturated. These single-electron basis states are required for the SD method and for constructing multielectron basis states in CI calculations.

In the SD method, the many-electron wave function $| \Psi \rangle$ is expanded in terms of the reference RHF state $| \Phi_0 \rangle$ and single and double excitations:
\begin{equation}
\begin{aligned}
|\Psi \rangle &= \left[ 1 + \sum_{ma} \rho_{ma} a_{m}^{\dagger} a_{a} + \frac{1}{2} \sum_{mnab} \rho_{mnab} a_{m}^{\dagger} a_{n}^{\dagger} a_{b} a_{a} \right. \\
& \quad \left. + \sum_{m \neq v} \rho_{mv} a_{m}^{\dagger} a_{\nu} + \sum_{mnb} \rho_{mnvb} a_{m}^{\dagger} a_{n}^{\dagger} a_{b} a_{\nu} \right] |\Phi_0\rangle,
\end{aligned}
\end{equation}
Indices $a$, $b$ refer to core states, indices $m$, $n$, $k$, $l$ refer to excited states, index $v$ refers to the valence state.
The coefficients $\rho_{ma}$, $\rho_{mnab}$,  $\rho_{mv}$ and $\rho_{mnvb}$ represent the expansion coefficients corresponding to the single- and double- electron excitations from the core and from the valence state. 
These coefficients are found by solving the SD equations (see, e.g.~\cite{CC1,CC2,CC3}).
Solving the SD equations gives the correlation operators $\hat \Sigma_1$ and $\hat \Sigma_2$ to be added to the CI Hamiltonian.
$\hat \Sigma_1$ is the one-electron operator that describes the correlation interaction of a particular valence electron with the core.
$\hat \Sigma_2$ is the two-electron operator that describes the screening of the Coulomb interaction between two valence electrons by core electrons. 
The operators are defined by
\begin{eqnarray}
&&\langle m|\hat \Sigma_1| v \rangle  = \sum_{bn}\tilde g_{mban}\rho_{nb} + 
 \nonumber \\
&&\sum_{bnr}g_{mbnr}\tilde\rho_{nrvb}- \sum_{bcn}g_{bcvn}\tilde\rho_{mnbc},
\label{e:S1me}
\end{eqnarray}
and
\begin{eqnarray}
&&\langle mn |\hat \Sigma_2| vw \rangle =  \nonumber \\
&&\sum_{cd}g_{cdvw}\rho_{mncd} -  \sum_c \left( g_{cnvw}\rho_{mc} 
+ g_{cmwv}\rho_{nc} \right) + \nonumber \\
&&\sum_{rc} \left( g_{cnrw}\tilde \rho_{mrvc} 
+ g_{cmrv}\tilde \rho_{nrwc}  + g_{cnwr} \rho_{mrvc}
\right. \label{qscreen} \\   
&&+ g_{cmvr} \rho_{nrwc} - g_{cmwr} \rho_{nrcv} - \left. g_{cnvr}
  \rho_{mrcw} \right).  \nonumber  
\end{eqnarray}

The tilde above $g$ or $\rho$ means the sum of direct and
exchange terms, e.g.
\[ \tilde \rho_{nrbc} = \rho_{nrbc} - \rho_{nrcb}. \]

Including the $\hat \Sigma_1$ and $\hat \Sigma_2$ operators, the CI Hamiltonian takes the form
\begin{equation}
    \hat{H}^{\mathrm{CI}} = \sum_i^{M} \left(\hat{H}^{\mathrm{RHF}}+ \hat{\Sigma}_1 \right)_i + \sum_{i<j}^{M} \left( \frac{e^2}{|\boldsymbol{r}_i-\boldsymbol{r}_j|} + \hat{\Sigma}_{2ij} \right),
    \label{HCI}
\end{equation}
where $i$ and $j$ are the electron indices. 
In principle, this Hamiltonian can be used to perform the calculations using the standard CI technique. However, to increase the efficiency of the calculations, we perform an additional step. We use the CIPT method~\cite{CIPT} (configuration interaction with perturbation theory). 
In this approach, the expansion of the many-electron wave function for a state $m$ is divided into two groups of terms:
\begin{equation}
|\Psi_m\rangle = \sum_{i \in P} c_{im} |\phi_i\rangle + \sum_{j \in Q} c_{jm} |\phi_j\rangle,
\end{equation}
{where $\phi_i$ are single-determinant many-electron basis functions constructed from a reference configuration (or a set of reference configurations) by exciting one or two electrons. Up to fourteen single-electron states above the core are used in each partial wave to construct basis states $\phi_i$.}
$P$ is the set of low-energy basis states treated explicitly in the CI matrix and $Q$ is set of high-energy states included perturbatively. 
To reduce the size of the CI matrix, all high-energy off-diagonal elements are neglected, only the diagonal along with the off-diagonal elements between $P$ and $Q$ are retained. We take $N_\mathrm{eff}$ as the number of states in the low-energy basis set $P$. The expansion coefficients $c_{im}$, $c_{jm}$ and corresponding energies $E_m$ are found by solving the eigenvalue equation with effective CI matrix of size $N_\mathrm{eff}$:
\begin{equation}
    (\hat{H}^{\text{eff}} - EI)X=0,
    \label{eigeneq}
\end{equation}
where $I$ is the unit matrix, $X = ( c_1, \dots, c_{N_\mathrm{eff}})^\mathrm{T}$. 
The effective CI matrix is given as the low-energy CI Hamiltonian with high-energy perturbative corrections:
\begin{equation}
H_{ij}^\text{eff} = H_{ij} + \sum_{k \in Q} \frac{H_{ik} H_{kj}}{E - E_k},
\label{e:e2}
\end{equation}
where $H_{ij}$ is the low-energy CI Hamiltonian with states $i,j \leq N_\mathrm{eff}$, $H_{ik}$ represents the matrix elements between states in $P$ and $Q$ with $N_\mathrm{eff} < k \leq N_s$, $E_k = \langle \phi_k | \hat{H}^\mathrm{CI} | \phi_k \rangle$ and $E$ is energy of the state of interest. Since $E$ is not known in advance, one has to perform iterations of Eqs.~(\ref{eigeneq}) and (\ref{e:e2}) until convergence in $E$ is achieved. 

To account for magnetic interactions and retardation effects, we include the Breit interaction in the calculations. Quantum electrodynamic (QED) radiative corrections are also incorporated using a radiative potential that accounts for both vacuum polarization and self-energy contributions ~\cite{radpot}. Details are described in our earlier works (see, e.g.,\cite{DzuEtAl01a,DzuFlaSaf06,Etotal}).
As both the Breit and QED radiative corrections scale with nuclear charge Z faster than the first power, their contribution to the energy levels of SHEs could be significant.
However, our calculations show that even for the superheavy elements Lv and Ts, the combined contribution of the Breit interaction and QED corrections {into excitation energies} does not exceed 100~cm$^{-1}$, which is smaller than the uncertainty caused by the correlations. This is partly due to the relaxation effect~\cite{QED-relaxation}, 
i.e. the modification of the RHF potential due to Breit and QED corrections altering   atomic electron wave functions. The relaxation effect significantly reduces the corresponding contributions. However, the Breit and QED corrections are included in all calculations just to make sure that nothing important is missed.

For each level, we calculate the Land\'{e} $g$-factor and compare it to both the experimental values and the non-relativistic expression
\begin{equation}\label{e:gf}
g = 1 + \frac{J(J+1)-L(L+1)+S(S+1)}{2J(J+1)}.
\end{equation}
Here, $J$ is the total angular momentum of all atomic electrons, $L$ is the angular momentum, and $S$ is the spin.
The calculated $g$-factor is obtained as the expectation value of the operator describing the interaction between the electrons and the magnetic field (M1): {$\hat H_{M1} = \overrightarrow{\mu} \cdot \overrightarrow{B}$, where $\overrightarrow{\mu}$ is the  pseudo-vector of atomic magnetic moment. In the relativistic case the matrix elements of this operator over single-electron wave functions
\begin {equation}
\psi(r)=\frac{1}{r}\left(\begin{array}{c} f(r)\Omega_{\kappa m} \\ i\alpha g(r) \Omega_{-\kappa m} \end{array} \right)
\end{equation}
are given by 
\begin{eqnarray}
\langle \psi_a|\hat H_{M1}|\psi_b\rangle &=& (\kappa_a+\kappa_b)\langle -\kappa_a||C^1||\kappa_b \rangle \nonumber \\
&\times&\int 2(f_ag_b+g_af_b)j_1(kr)dr B,
\end{eqnarray}
where $\kappa = l$ for $j=l-1/2$, $\kappa = -l-1$ for $j=l+1/2$, $C^1$ is the normalized spherical harmonic, wave vector $k=\omega/c$, 
$J_1(kr)$ is the spherical Bessel function.
The $g$-factor is defined via energy shift $\Delta E = \mu_B g M_J B$, where $M_J$ is the projection of the total atomic angular momentum $J$.
For many-electron wave function $v$ for valence electrons we have
\begin{equation}\label{e:gv}
g_v = \frac{1}{\sqrt{J_v(2J_v+1)(J_v+1)}} \langle v|| \hat H_{M1}||v\rangle/B.
\end{equation}
}

Since $L$ and $S$ are not well defined in the  relativistic calculations, we use values of $L$ and $S$ that bring (\ref{e:gf}) into the best possible agreement with the calculated $g$-factors. Comparing with the non-relativistic expression helps in identifying and naming the states, as well as grouping them into fine structure multiplets.

The ionization potential is calculated as the difference between the ground state energy of the neutral atom and the ground state of the positive ion. Similarly, the electron affinity is calculated as the difference between the ground state energy of the neutral atom and the ground state of the negative ion.

\section{Energy levels of Te and I.}

\begin{table}[ht]
\caption{Calculated energy levels and $g$-factors of tellurium, compared with experiment. $\Delta$ is the difference between the calculated value and NIST value of energy.}
\label{tab:te}
\begin{tabular}{cccccccccc}
\hline
\hline

 &  &  &  & \multicolumn{3}{c}{Energy   ($\text{cm}^{-1}$)} &  & \multicolumn{2}{c}{Land\'{e} $g$-factor} \\ \cline{5-7} \cline{9-10} 
Conf. & Term & J &  & This & NIST \cite{NIST_ASD} & $\Delta$ &  & This & NIST \cite{NIST_ASD} \\ 
 &  &  &  & work &  &  &  & work &  \\ \hline
$5p^4$ & ${}^3\text{P}$ & 2 &  & 0 & 0 &  &  & 1.4655 & 1.460(4) \\
                     &  & 0 &  & 4568 & 4706 & -139 &  & 0 &  \\
                     &  & 1 &  & 4834 & 4750 &   83 &  & 1.5001 & 1.514(10) \\
$5p^4$ & ${}^1\text{D}$ & 2 &  & 10719 & 10557 & 161 &  & 1.0347 & 1.037(3) \\
$5p^4$ & ${}^1\text{S}$ & 0 &  & 22932 & 23198 & -266 &  & 0 &  \\

$5p^36s$ & ${}^5\text{S}^{\circ}$ & 2 &  & 43235 & 44253 & -1018 &  & 1.9658 & 1.960(3) \\ %
$5p^36s$ & ${}^3\text{S}^{\circ}$ & 1 &  & 45709 & 46652 & -944 &  & 1.9179 & 1.909(3) \\ %

$5p^36p$ & ${}^5\text{P}$ & 1 &  & 54388 & 54160 & 228 &  & 2.4327 & 2.388(2) \\  
                       &  & 2 &  & 54373 & 54199 & 174 &  & 1.8039 & 1.762(1) \\ 
                       &  & 3 &  & 54377 & 54535 & -158 &  & 1.6472 & 1.636(3) \\ 

$5p^36s$ & ${}^3\text{D}^{\circ}$ & 1 &  & 54289 & 54683 & -395 &  & 0.6570 & 0.674(5) \\ 
                               &  & 2 &  & 54314 & 54880 & -566 &  & 1.2249 & 1.224(5) \\ 
                               &  & 3 &  & 56041 & 56842 & -802 &  & 1.3973 & 1.352(5) \\ 

$5p^36p$ & ${}^3\text{P}$ & 1 &  & 55371 & 55355 & 15 &  & 1.4850 & 1.504(3) \\ 
                       &  & 2 &  & 55793 & 55667 & 125 &  & 1.4944 & 1.496(5) \\ 
                       &  & 0 &  & 55830 & 55809 & 21 &  & 0 &  \\  

$5p^35d$ & ${}^5\text{D}^{\circ}$ & 3 &  & 55549 & 55677 & -129 &  & 1.4733 &  \\ %
                               &  & 4 &  & 55679 & 55816 & -137 &  & 1.4855 & 1.47(2) \\ %
                               &  & 2 &  & 55713 & 55817 & -104 &  & 1.4497 & 1.428(14) \\ %
                               &  & 0 &  & 55676 & 55826 & -151 &  & 0 &  \\   %
                               &  & 1 &  & 55713 & 55851 & -138 &  & 1.4850 &  \\ %

$5p^36s$ & ${}^1\text{D}^{\circ}$ & 2 &  & 56551 & 57114 & -563 &  & 1.0856 & 1.08(3) \\ %

$5p^35d$ & ${}^3\text{D}^{\circ}$ & 2 &  & 58321 & 58592 & -273 &  & 1.1368 & 1.112(1) \\ %
                               &  & 1 &  & 58578 & 58746 & -168 &  & 0.5274 & 0.532(4) \\ %
                               &  & 3 &  & 58555 & 58826 & -272 &  & 1.3162 & 1.310(10) \\ %
\hline \hline 
 
\end{tabular}

\end{table}

\begin{table}[ht]
\caption{Calculated energy levels and $g$-factors of iodine, compared with experiment.}
\label{tab:i}
\begin{tabular}{cccccccccc}
\hline
\hline
 &  &  &  & \multicolumn{3}{c}{Energy ($\text{cm}^{-1}$)} &  & \multicolumn{2}{c}{Land\'{e} $g$-factor} \\ \cline{5-7} \cline{9-10} 
Conf. & Term & J &  & This & NIST \cite{NIST_ASD} & $\Delta$ &  & This & NIST \cite{NIST_ASD} \\ 
 &  &  &  & work &  &  &  & work & \\ \hline

$5p^5$ & ${}^2\text{P}^{\circ}$ & 3/2 &  & 0 & 0 &  &  & 1.3333 &  \\
                             &  & 1/2 &  &  7664 & 7602  &    61 &  & 0.6667 & 0.673 \\
           $5p^46s$ & ${}^2[2]$ & 5/2 &  & 53193 & 54633 & -1440 &  & 1.5742 & 1.576 \\
                             &  & 3/2 &  & 54885 & 56092 & -1208 &  & 1.4041 & 1.385 \\
           $5p^46s$ & ${}^2[0]$ & 1/2 &  & 60023 & 60896 & -973  &  & 2.6209 & 2.561 \\
           $5p^46s$ & ${}^2[1]$ & 3/2 &  & 60250 & 61819 & -1570 &  & 1.6145 & 1.618 \\
                             &  & 1/2 &  & 62117 & 63186 & -1070 &  & 0.7198 & 0.799 \\
   $5p^46p$ & ${}^2[2]^{\circ}$ & 5/2 &  & 65744 & 64906 &   838 &  & 1.4752 & 1.524 \\
                             &  & 3/2 &  & 65813 & 64989 &   823 &  & 1.6173 & 1.619 \\
   $5p^46p$ & ${}^2[3]^{\circ}$ & 5/2 &  & 66722 & 65644 &  1078 &  & 1.2651 & 1.217 \\
                             &  & 7/2 &  & 66788 & 65669 &  1118 &  & 1.4078 & 1.42 \\
  $5p^46p$ & ${}^2[1]^{\circ}$  & 1/2 &  & 67306 & 65856 &  1449 &  & 1.6047 & 1.556 \\
                             &  & 3/2 &  & 68637 & 67061 &  1575 &  & 1.4060 & 1.415 \\ \hline \hline
			     
\end{tabular}
\end{table}

We start the calculations from the energy levels and the $g$-factors of tellurium and iodine. These are the heaviest atoms which have electron structures similar to those of Lv and Ts for which sufficient experimental data are available. Comparing calculated and experimental data helps to get an idea about the accuracy of the calculations. The results for Te and I are presented in Tables~\ref{tab:te} and \ref{tab:i}. 

Comparison with experiment shows that the average absolute difference between theory and experiment($|\Delta|$) is $\sim 300$~cm$^{-1}$ for tellurium, which is $< 1$\% of the excitation energy. The corresponding numbers for iodine are $\sim 1000$~cm$^{-1}$ and $\sim$~2\%.
The slightly better accuracy for Te is most likely due to the smaller number of valence electrons. 
{This is another important feature of the $V^{N-M}$ approximation used in the present work. If the $V^{N}$ approximation were used instead—where all valence electrons are included in the initial RHF procedure—one might expect better accuracy for systems with a smaller number of holes in the open subshell. That is, the accuracy for iodine would likely be better than for tellurium. In the case of the $V^{N-M}$ approximation, the initial approximation is less accurate for atoms with  a larger number of valence electrons. This may reflect on the final accuracy.  However, the  final results for both atoms would probably  be less accurate in $V^{N}$ approximation compared to those presented here, due to the omission of the core-valence correlations, which are difficult to include in the $V^{N}$ approximation (see previous section for discussion).

In our experience, to account for the core-valence correlations is  more important than to start from a good initial approximation. On the other hand, a complete basis is required to compensate for the poorer initial approximation. Achieving the basis set saturation becomes increasingly difficult as the number of valence electrons grows.

The accuracy of the calculations for heavier atoms—polonium, astatine, livermorium, and tennessine—is expected to be similar to that found for tellurium and iodine in this section.
}


\section{Energy levels of Po and At.}

\begin{table}[ht]
\caption{Calculated energy levels and $g$-factors of polonium, compared with experiment.}
\label{tab:po}
\begin{tabular}{ccc ccc c}
\hline
\hline

&  &  &   \multicolumn{3}{c}{Energy   ($\text{cm}^{-1}$)} &   \multicolumn{1}{c}{$g$-factor} \\ \cline{4-6} 
Conf. & Term & J &  Calc. & NIST \cite{NIST_ASD} & $\Delta$ &  Calc.  \\ 
\hline \hline
$6p^4$   & $^3\text{P}$           & 2 &     0 &  0       & 0 & 1.3995  \\
         &                        & 0 &  7072 &  7515 & -443 &   0   \\
         &                        & 1 & 16381 & 16832 & -451 & 1.5007  \\
$6p^4$ & $^1\text{D}$             & 2 & 21310 & 21679 & -369 & 1.1009 \\
$6p^4$ & $^1\text{S}$             & 0 & 43210 & 42718 &  492 &  0  \\
$6p^37s$ & ${}^5\text{S}^{\circ}$ & 2 & 38816 & 39081 & -265 & 1.7169  \\
$6p^37s$ & $^3\text{S}^{\circ}$   & 1 & 40721 & 40803 &  -82 & 1.4169 \\							 
$6p^37p$ & $^5\text{P}$           & 1 & 50506 &         & & 1.6218 \\
         &                        & 2 & 50614 &         & & 1.4532 \\
         &                        & 3 & 51667 & 50681\footnotemark[1] & 986 & 1.4807  \\	
$6p^37s$ & $^3\text{D}^{\circ}$   & 1 & 53005 & 52532   &  473 & 1.2731 \\
         &                        & 2 & 52163 & 51713   &  450 & 1.1196  \\
         &                        & 3 & 52591 & 52098   &  493 & 1.3257 \\
	 
$6p^37p$ & $^3\text{P}$           & 1 & 52003 &         & & 1.7028  \\ 
         &                        & 2 & 53152 &         & & 1.3977  \\ 
         &                        & 0 & 54068 &         & & 0   \\ 							 
$6p^36d$ & $^5\text{D}^{\circ}$   & 3 & 56130 &         & & 1.2108   \\   
         &                        & 4 & 52834 &         & & 1.3593 \\
         &                        & 2 & 53642 & 53028   & 584 & 1.3144 \\
         &                        & 0 & 53253 &         & & 0 \\
         &                        & 1 & 54611 & 53762   & 849 & 1.0545 \\
	 

$6p^36d$ & $^3\text{D}^{\circ}$   & 2 &  55058 & 55465  & -407 & 1.7272  \\
         &                        & 1 &  54611 & 54250  &  361 & 0.8524  \\
         &                        & 3 &  55045 &        & & 1.3255 \\
\hline \hline 
\end{tabular}
\footnotetext[1]{The value $J=3$ is under question in NIST database~\cite{NIST_ASD}. If we instead assign $J=2$, there could be perfect agreement with the calculated level 50614 cm$^{-1}$.}
\end{table}

\begin{table}[ht]
\caption{Calculated energy levels and $g$-factors of astatine, compared with experiment}
\label{tab:at}
\begin{tabular}{ccc ccc c}
\hline
\hline
 &  &  &   \multicolumn{3}{c}{Energy   ($\text{cm}^{-1}$)} &   \multicolumn{1}{c}{$g$-factor} \\ \cline{4-6} 
Conf. & Term & J &  Calc. & NIST \cite{NIST_ASD} & $\Delta$ &  Calc.  \\ 
\hline
\hline
$6p^5$ & $^2\text{P}^{\circ}$ & 3/2 &     0 & 0 & 0 &1.3337   \\
              &               & 1/2 & 22547 &  &  & 0.6667  \\            
$6p^47s$ & $^2[2]$            & 5/2 & 44590 & 44549 & 41 & 1.5178 \\      
         &                    & 3/2 & 46686 & 46234 & 362 & 1.2735  \\    
$6p^47s$ & ${}^2[0]$          & 1/2 & 54373 &        && 2.1089 \\         
$6p^47s$ & ${}^2[1]$          & 3/2 & 59533 &        && 1.1995 \\      
         &                    & 1/2 & 59896 &        && 0.8925 \\      
         
$6p^47p$ & ${}^2[2]^{\circ}$ & 5/2 & 59342 &        && 1.3924  \\   
         &                   & 3/2 & 59229 &        && 1.5455  \\   
         
$6p^47p$ & ${}^2[3]^{\circ}$ & 5/2 & 59585 &        && 1.2983 \\  
         &                   & 7/2 & 59441 &        && 1.3289  \\ 
         
$6p^47p$ & ${}^2[1]^{\circ}$ & 1/2 & 60184 &        && 1.4071  \\  
         &                   & 3/2 & 62610 &        && 1.3742  \\ 
         
$6p^46d$ & ${}^2[3]$         & 7/2 & 60060 &        && 1.3680   \\  
         &                   & 5/2 & 60011 &        && 1.2386  \\  
	 
$6p^46d$ & ${}^2[1]$         & 3/2 & 62439 &        && 1.2705   \\ 
         &                   & 1/2 & 62430 &        && 1.9364   \\
\hline \hline 
\end{tabular}
\end{table}

Calculations for Po, I, and At play an intermediate role. On the one hand, the available experimental data help assess the accuracy of the calculations and verify the conclusions and assumptions from the previous section. On the other hand, since the experimental data are very limited, the calculations can help fill in the gaps in the data.

The results of the calculations are presented in Tables \ref{tab:po} and \ref{tab:at}.
Comparison with experimental data for Po shows a similar trend to that of Te. Note that experimental data for Po are limited and that some highly excited states are difficult to identify. No $g$-factors are known and, for some states, even the values of the total angular momentum $J$ are not known. For example, for the state with $E$=50681~cm$^{-1}$, the value of $J=3$ is brought under question in the NIST database~\cite{NIST_ASD}. If $J=2$ is assumed instead, then the agreement with theory is excellent. 

Only two energy levels of At I are presented in the NIST database. For both levels, the agreement between theory and experiment is very good (see Table~\ref{tab:at}).

The difference in spectra between heavy many-valence-electrons elements and their lighter analogs is the result of the complicated interplay between relativistic and correlation effects. Leading relativistic effects are proportional to $(Z\alpha)^2$ ($\alpha \approx 1/137$ is the fine structure constant), which ranges from 0.14 (Te) to 0.73 (Ts) for atoms considered in the present work. Relativistic effects shift the energies, which in turn affect configuration mixing since the mixing is inversely proportional to the energy interval between the mixing states. 

To see how large relativistic energy shifts are, it is instructive to consider the shifts of the single-electron energies \cite{DzuFlaWeb99}
\begin{equation} \label{e:deltaen}
\Delta E_n = \frac{E_n}{\nu} (Z\alpha)^2 \left( \frac{1}{j+1/2} -0.6 \right).
\end{equation}
Here $\nu$ is the effective principal quantum number ($E_n = -1/(2\nu^2)$), $j$ is the total angular momentum of the single-electron state.
This formula was obtained from the consideration of hydrogen-like systems. The parameter -0.6 was introduced to fit many-body effects,
such as the Hartree-Fock exchange interaction. Given that for valence states $\nu \sim 1$ we see that the correction is about 6\% of the single-electron energy of the $s_{1/2}$ and $p_{1/2}$-states of Te and I, $\sim$ 15\% for Po and At, and $\sim$ 30\% for Lv and Ts.  
It is useful to keep in mind that the correction is maximal in value and negative for $j=1/2$ ($E_n$ is negative), while it is smaller and positive for all states with $j>1/2$.

Large relativistic effects cause significant differences in the relative positions on the energy scale of atomic states with different electronic configurations. However, the difference between the calculated and experimental energy levels is very similar for Te, I, Po, and At. 
We expect similar accuracy for heavier elements Lv and Ts.

\section{Energy Levels of Lv I and Ts I}

\begin{table}[ht]
\caption{Calculated energy levels and $g$-factors of livermorium.}
\label{tab:lv}
\begin{tabular}{ccccccc}
\hline
\hline

Configuration & Term & J &  & Energy ($\text{cm}^{-1}$)  & $g$-factor \\ \hline
$7p^4$ & ${}^3\text{P}$ & 2 &  & 0 &     1.3573 \\
                      & & 0 &  & 6605 &    0 \\ 
                      & & 1 &  & 37319 &  1.4309 \\
		      
$7p^4$ & ${}^1\text{D}$ & 2 &  & 37505 &  1.2330 \\
$7p^4$ & ${}^1\text{S}$ & 0 &  & 43184 &  0 \\

$7p^38s$ & ${}^5\text{S}^{\circ}$ & 2 &  &   24790 &  1.5517 \\
$7p^38s$ & ${}^3\text{S}^{\circ}$ & 1 &  &   27190 &  1.2096 \\

$7p^38p$ & ${}^5\text{P}$         & 1 &  &   40675 &  1.5120 \\
                                                     &  & 2 &  &   41959 &  1.3418 \\
                                                     &  & 3 &  &   40628 &  1.3650 \\
						     
$7p^38s$ & ${}^3\text{D}^{\circ}$ & 1 &  &   43478 &  1.0511 \\
                                                     &  & 2 &  &   40819 &  0.9909 \\
                                                     &  & 3 &  &   41467 &  1.2195 \\
						     
$7p^38p$ & ${}^3\text{P}$         & 1 &  &   48371 &  1.4850 \\
                                                     &  & 2 &  &   48394 &  1.2345 \\
                                                     &  & 0 &  &   49916 &  0 \\
						     
$7p^37d$ & ${}^5\text{D}^{\circ}$ & 3 &  &   43322 &  1.1302 \\
                                                     &  & 4 &  &   41499 &  1.2724 \\
                                                     &  & 2 &  &   42351 &  1.3272 \\
                                                     &  & 0 &  &   42319 &  0 \\
                                                     &  & 1 &  &   42181 &  1.1779 \\
						     
\hline \hline 
\end{tabular}
\end{table}

\begin{table}[ht]
\caption{Calculated energy levels and $g$-factors of tennessine. Previous multiconfigurational Dirac-Fock (MCDF) calculations are included.}
\label{tab:ts}
\begin{tabular}{ccccc }
\hline
\hline
Configuration & Term & J &   Energy ($\text{cm}^{-1}$)  & $g$-factor \\
\hline
$7p^5$ & ${}^2\text{P}^{\circ}$      & 3/2 &   0 & 1.3358 \\
                                  &  & 1/2 &   46920 &  1.3443 \\ 
				  
$7p^48s$ & ${}^2[2]$ & 5/2 &   27658\footnotemark[1] & 1.4859 \\  
                                  &  & 3/2 &   31208 & 1.2241 \\   
				  
$7p^48s$ & ${}^2[0]$ & 1/2 &   38506 &  2.0238 \\ 
$7p^48s$ & ${}^2[1]$ & 3/2 &   51827\footnotemark[2] & 1.2394 \\ 
                                  &  & 1/2 &   47110 & 0.8100 \\ 
				  
$7p^48p$ & ${}^2[2]^{\circ}$
                                   &   5/2  &   44091 & 1.2221 \\   
                                  &  & 3/2 &   43889 & 1.4889 \\ 
$7p^48p$ & ${}^2[3]^{\circ}$
                                   &    5/2 &   47032 & 1.3456 \\ 
                                  &  & 7/2 &   47058 & 1.3458 \\ 
$7p^48p$ & ${}^2[1]^{\circ}$
                                   &    1/2 &   46920 &  1.3443  \\
                                  &  & 3/2 &   49625 &  1.3546  \\
				  	     	     	         
$7p^47d$ & ${}^2[3]$ & 7/2 &   46789 &  1.2948  \\
                                  &  & 5/2 &   47186 &  1.1654  \\
$7p^47d$ & ${}^2[1]$ & 3/2 &   47146 &  1.1654  \\
                                  &  & 1/2 &   49193 &  2.0802  \\
				  \hline \hline 
\end{tabular}
\footnotetext[1]{MCDF, $E$=26116 cm$^{-1}$~\cite{Z117}.}
\footnotetext[2]{MCDF, $E$=51958 cm$^{-1}$~\cite{Z117}.}
\end{table}

The calculated energy levels of Lv and Ts are presented in Tables \ref{tab:lv} and \ref{tab:ts}. Experimental data for these superheavy atoms are absent.
Calculations using the multi-configurational Dirac-Fock (MCDF) method were previously performed for Ts~\cite{Z117}. The agreement between our results and those from the MCDF method is good and consistent with the accuracy estimates discussed in previous sections.

The spectra of Lv and Ts show the continuation of the trend observed in the comparison of the spectra of Te and I with Po and At.
The trend is the reduction of the energy interval between states of different configurations. This is a fortunate feature because it brings more electric dipole transitions into the optical region. There are at least five electric dipole transitions between ground and excited states of Lv which are in the optical region.
There are at least three such transitions in Ts. These are $7p$ - $8s$ transitions in the single-electron approximation, which suggests that they should be strong. Note that the change of spin in the transition would normally lead to suppression in the non-relativistic limit. However, the relativistic effects in Lv and Ts are sufficiently strong to eliminate this suppression.

Note that the $LS$-coupling scheme still works well for the superheavy elements Lv and Ts. At least the states can be grouped into fine-structure multiplets. The deviation from the scheme is significant, which is evident from the values of the $g$-factors. For example, the $g$-factors for the ground-state $^3$P multiplet of Lv in the non-relativistic limit should be $g$=1.5 for $J$=1 and 2, and  $g$=0 for $J$=0 (see Eq.~(\ref{e:gf})).
In fact, $g$=1.43 for $J=1$ and $g$=1.36 for $J$=2 (see Table~\ref{tab:lv}). A similar picture takes place for other multiplets of both atoms.
Making the same fine-structure multiplets and the same state labels for heavy and light atoms makes it easier to compare the spectra.

\subsection{Electric dipole amplitudes and field isotope shift}

We consider now electric dipole amplitudes and field isotope shift for optical E1 transitions in superheavy elements Lv and Ts.

Calculations start from solving the RPA equations (random-phase approximation or time-dependent Hartree-Fock method; see, e.g. \cite{DzuFlaSilSus87}). 
\begin{equation}\label{e:RPA}
\left(\hat H_0 - \epsilon_c\right) \delta \psi_c = -\left(\hat F +\delta V^{F}\right) \psi_c.
\end{equation}  
Here, $\hat H_0$ is the relativistic Hartree-Fock operator for the closed-shell core.
$\hat F$ is the operator of external field. In the case for electric dipole amplitudes $\hat F = -er$; 
in the case of FIS, the operator  $\hat F = \delta V_N$ describes the change in nuclear potential due to the change in nuclear radius. 
The index $c$ numerates the core states, $\psi_c$ and $\delta \psi_c$ are single-electron states of the core and corresponding correction to it due to external electric field; 
$\delta V^{F}$ is the correction to the self-consistent Hartree-Fock potential $V^{N-M}$ caused by the change $\delta \psi_c$ in every core state $c$.

Solving the RPA equations (\ref{e:RPA}) determines the effective operator of the external field. The amplitude of the electric dipole transition between many-electron states $a$ and $b$ is given by
\begin{equation} \label{e:A}
A_{ab} = \langle a||\mathbf{d}+\delta V^{d}||b\rangle.
\end{equation}
The FIS between the same states is given by
\begin{eqnarray} \label{e:FIS}
&\Delta E_{ab} = \langle a | \delta V_N + \delta V^{F} | a \rangle -  \langle b | \delta V_N + \delta V^{F} | b \rangle \\
 &\equiv F_{ab} \delta \langle r^2 \rangle. \nonumber
\end{eqnarray}
The results of calculations for $A_{ab}$ and $F_{ab}$ are presented in Table~\ref{t:E1}.
We see in particular that the E1 amplitudes are not suppressed in spite of the change of spin in the $LS$-coupling scheme.
This is due to strong relativistic effects. The values of the FIS constants are also relatively large due to the change in the number of $s_{1/2}$ or $p_{1/2}$ electrons in the transitions.

\begin{table} 
\caption{Amplitudes of optical electric dipole (E1) transitions and field isotope shifts (FS) 
for some transitions between ground ($a$) and excited ($b$) states  of Lv and Ts.}
\label{t:E1}
\begin{ruledtabular}
\begin{tabular}{ccccc}
Atom & \multicolumn{2}{c}{Upper state} & $\langle a||E1||b\rangle$ & $F_{ba}$ \\
     & Term & $E$ (cm$^{-1}$) & (a.u.) & (GHz/fm$^2$) \\
\hline
Lv & $^5$S$^o_2$ & 24790 & 3.39 & 295 \\
   & $^3$S$^o_1$ & 27190 & 3.26 & 261 \\
   & $^3$D$^o_1$ & 43478 & 1.10 & 169 \\
   & $^3$D$^o_2$ & 40819 & 0.32 & 166 \\
   & $^3$D$^o_3$ & 41467 & 1.97 & 167 \\

Ts & $^2[2]_{5/2}$ & 27658 & 1.62 & 341 \\
   & $^2[2]_{3/2}$ & 31208 & 2.98 & 291 \\
   & $^2[0]_{1/2}$ & 38506 & 2.26 & 327 \\
\end{tabular}
\end{ruledtabular}
\end{table}

\section{Ionization Potential and Electron Affinity}

\begin{table} 
\caption{Ionization potentials (in cm$^{-1}$) of Te, I, Po, At, Lv and Ts.
$\Delta$ is the difference between present calculations and experiment.}
\label{t:IPs}
\begin{ruledtabular}
\begin{tabular}{cccrl}
Atom & Present & Expt. & \multicolumn{1}{c}{$\Delta$} & Other calculations \\
\hline
Te &  71661 & 72669 \cite{KIECK}    & -1008  &  \\
Po &  67695 & 67896 \cite{FINK201972} &  -201 & \\
Lv &  54433 &         &  & 55290 \cite{Borschevsky_2015}, 59600 \cite{Z116},  \\
   &        &      &    &  53470 \cite{Dyall2012}   \\
   &        &     &      &    \\
I  &  82612 & 84295 \cite{Esteves_2023} &  -1683 & \\
At &  74639 & 75151 \cite{Rothe2013}    & -512 &  \\
Ts &  61643 &        &   &   61730 \cite{Borschevsky_2015}, 59520 \cite{Dyall2012}, \\
&&&                  &   61600 \cite{Z117}, 61310 \cite{Liu_2006},  \\
&&&                  &   61100 \cite{Mitin} \\ 
\end{tabular}
\end{ruledtabular}
\end{table}

\begin{table} 
\caption{Electron affinities (in cm$^{-1}$) of Te, I, Po, At, Lv and Ts.
$\Delta$ is the difference between present calculations and experiment.}
\label{t:EAs}
\begin{ruledtabular}
\begin{tabular}{cccrl}
Atom & Present & Expt. & \multicolumn{1}{c}{$\Delta$}  & Other calculations \\
\hline
Te & 14970 & 15896 \cite{Haeffler1996} & -926  \\
Po & 12002 &         &  & 11850  \cite{Borschevsky_2015}, 11331 \cite{Li_2012}   \\
Lv &  5623 &         &  & 6260 \cite{Borschevsky_2015}, 5000 \cite{Dyall2012} \\
   &  &  &  &  \\
I  & 23169 & 24673 \cite{Pelaez_2009} &  -1504 & \\
At & 19710 & 19485 \cite{Leimbach2020} &  225 &  \\
Ts & 11579 &         &  & 12920 \cite{Borschevsky_2015}, 11320 \cite{Dyall2012},   \\
          &&&&                11700  \cite{Z117}, 11040 \cite{Thierfelder}, \\
   &&    &                &  12860 \cite{Liu_2006}, 11900 \cite{Mitin} \\  
\end{tabular}
\end{ruledtabular}
\end{table}

Ionization potential (IP) and electron affinity (EA) are important atomic characteristics closely related to their chemical properties.
Experimental data are limited, especially for Lv and Ts. However, there is a large number of theoretical works that calculate IP and EA of superheavy elements and their lighter analogs using a wide range of methods with different levels of sophistication. 
A detailed review of these works goes beyond the scope of the present article. Good review papers with many references can be found, e.g. in Refs.~\cite{smits2023,smits2024,Borschevsky_2015}. Here we focus on superheavy elements Lv and Ts using their lighter analogs mostly to illustrate the accuracy of the calculations.

Calculations of the IP and EA are very similar to those of the spectra. One just needs to change the number of electrons in the CI
calculations. For example, the IP is found as a difference between ground state energies of the neutral atom and its positive ion. 
Similarly, EA is the difference in the ground state energies of a neutral atom and its negative ion. This similarity means in particular that the agreement for IPs and EAs between theory and experiment and with other calculations is another indication of the accuracy of the spectra calculations.

Our results for IPs and EAs are presented in Tables \ref{t:IPs} and \ref{t:EAs} and compared with available experimental data or with other calculations when experimental data are absent. Note that the agreement between our calculated IPs and experiment is between 1 and 2\%.
The agreement between our results for Lv and Ts and the most advanced calculations of Ref.~\cite{Borschevsky_2015} is on the same level.
It is natural to assume that the IPs of Lv and Ts are now known to an accuracy of $\sim$~1\%.

The difference between different results for EAs is greater. 
{This can be partly explained by the fact that negative ions are weakly bound systems, for which basis set saturation is difficult to achieve.}
Note that the difference between ground state energies of a neutral atom and its negative ions is only $\sim$~0.5\% of the binding energy of all valence electrons. This means that if the numerical error is higher for a negative ion, then the error is enhanced in the difference. In the end, we may assume that the EAs of Lv and Ts are now known to an accuracy of $\sim$~10\%.

\section{Static dipole polarizabilities}

Static dipole polarizabilities ($\alpha_0$) are important characteristics of neutral atoms. They are related to the interactions of the atoms with the environment via a long-range potential $V_{pol}(r) = -\alpha_0/r^4$. The available data on the polarizabilities of superheavy elements are limited.
The most complete compilation of theoretical and experimental data on the polarizabilities of neutral atoms \cite{pol0} has no data on the Lv atom and only one reference to an estimation of the polarizability of Ts. Accurate calculations of the polarizabilities of the atoms considered in the present work is challenging due to the large number of valence electrons (see, e.g. \cite{Symmetry}). Fortunately, there is an approximate method that works well just for these types of atoms~\cite{DKF14}. 
The method is good  for the calculation of the static scalar dipole polarizabilities of  atoms with closed or nearly closed shells.

Note that all atoms considered in the present work have one or two holes in otherwise closed $p$-shell.
Therefore, they can be considered approximately as atoms with closed shells but with reduced contribution of the open $p$-shell into the inter-electron potential. The latter is done via fractional occupation numbers. Note that one should take care about canceling the self-action in the Hartree-Fock equations. This cancellation is exact for a closed-shell atom. Introduction of the fractional occupation numbers breaks this cancellation which needs to be restored "by hand". We will not consider these rather technical details further.

Calculations start from solving the RPA equations for the whole atom.
\begin{equation}\label{e:RPA1}
\left(\hat H_0 - \epsilon_c\right) \delta \psi_c = -\left(\mathbf{d}+\delta V^{d}\right) \psi_c.
\end{equation}  
These equations differ from (\ref{e:RPA}) by inter-electron potential. The $V^{N-M}$ potential for the closed-shell core is used in (\ref{e:RPA}) while here we use the potential $V$ which is calculated for all electrons including open shells and using fractional occupation numbers for them.
The index $c$ numerates the electron states, $\psi_c$ and $\delta \psi_c$ are single-electron wave functions and corresponding correction to them due to an external electric field; $\mathbf{d} = -e\mathbf{r}$ is the electric dipole operator; $\delta V^{d}$ is the correction to the self-consistent Hartree-Fock potential $V$ caused by the change $\delta \psi_c$ in every electron state $c$.

Equations (\ref{e:RPA}) are solved self-consistently for all atomic states.
Then static dipole polarizability is given by \cite{DKF14}
\begin{equation}\label{e:pol0}
\alpha_0 = \frac{2}{3}\sum_c \langle \delta \psi_c|| \mathbf{d}|| \psi_c \rangle.
\end{equation}
The approach based on (\ref{e:RPA1}) and (\ref{e:pol0}) should work well for systems which do not have low-energy electric dipole excitations from the ground state. The analysis of the spectra of the atoms of interest in previous sections shows that the best accuracy should be expected for Te and I, while it may slightly deteriorate for heavier atoms.

The results of the calculations are presented in Table~\ref{t:pol0}. The last column of the table presents recommended values from Ref.~\cite{pol0}.
For light atoms, the recommended values come from the analysis of a relatively large number of experimental and theoretical results.
The most accurate result is for the iodine atom. It is mainly based on the experimental measurements of Ref.~\cite{iodine-pol}. There are no data for Lv and there is only one result for Ts based on the estimation of Ref.~\cite{Ts-pol}. Comparing the results of our {\em ab initio} calculations with the recommended values of Ref.~\cite{pol0} one can conclude that the calculated values are probably underestimated by about 10\%.
Therefore, we can improve the predictions for heavy elements by multiplying all {\em ab initio} results by 1.1. 
The corresponding numbers are called "extrapolated" in Table~\ref{t:pol0}.

\begin{table} 
\caption{Static dipole polarizabilities (in atomic units) of Te, I, Po, At, Lv and Ts.}
\label{t:pol0}
\begin{ruledtabular}
\begin{tabular}{r cccc}
\multicolumn{1}{c}{$Z$} & Atom & \multicolumn{2}{c}{Present} & Ref.~\cite{pol0} \\
&& ab initio & Extrapolated & \\
\hline
52  & Te & 33.5 & & $38 \pm 4$ \\
84  & Po & 41.4 & 45.5 & $44 \pm 4$ \\
116 & Lv & 69.1 & 76.0 &\\
      & & & \\
53  & I   & 30.1 & &$ 32.9 \pm 1.3$ \\
85  & At & 38.2 & 42.0 & $ 42 \pm 4$ \\
117 & Ts & 65.1 & 71.6 &  $76 \pm 15$ \\
\end{tabular}
\end{ruledtabular}
\end{table}

\section{Conclusion}

We have calculated the electronic structure of the group 16 heavy elements Te, Po, and Lv, as well as group 17 heavy elements I, At, and Ts.
These include superheavy elements Lv ($Z$=116) and Ts ($Z$=117). Advanced theoretical techniques were used, such as the linearized coupled-cluster method and a combination of the configuration interaction method with perturbation theory. Energy levels, ionization potentials, electron affinities, and static dipole polarizabilities of the ground states of the atoms were calculated.
Calculations for heavier elements Po, At, Lv, and Ts address the gaps in the experimental data. Calculations for lighter elements, Te and I, demonstrate the accuracy of the calculations.
Comparison with available experimental data and other calculations indicate that the accuracy for energy levels, ionization potential, and electron affinities is on the level of a few percent. The accuracy for the polarizabilities is about 10\%.

\begin{acknowledgments}
This work was supported by the Australian Research Council Grant No. DP230101058.
\end{acknowledgments}
%

\end{document}